\documentclass[%
 reprint,
superscriptaddress,
 amsmath,amssymb,
 aps,
prb,
]{revtex4-2}

\usepackage{graphicx}
\usepackage{dcolumn}
\usepackage{bm}
\usepackage{xcolor}

\begin{document}

\preprint{APS/123-QED}

\title{Accelerating the Training and Improving the Reliability of Machine-Learned Interatomic Potentials for Strongly Anharmonic Materials through Active Learning}

\author{Kisung Kang}
\affiliation{ 
The NOMAD Laboratory at the FHI of the Max-Planck-Gesellschaft, Faradayweg 4-6, 14195 Berlin, Germany
}

\author{Thomas A. R. Purcell}
\affiliation{ 
The NOMAD Laboratory at the FHI of the Max-Planck-Gesellschaft, Faradayweg 4-6, 14195 Berlin, Germany
}

\author{Christian Carbogno}
 \email{carbogno@fhi-berlin.mpg.de}
\affiliation{ 
The NOMAD Laboratory at the FHI of the Max-Planck-Gesellschaft, Faradayweg 4-6, 14195 Berlin, Germany
}

\author{Matthias Scheffler}%
\affiliation{ 
The NOMAD Laboratory at the FHI of the Max-Planck-Gesellschaft, Faradayweg 4-6, 14195 Berlin, Germany
}

\date{\today}

\begin{abstract}
Molecular dynamics (MD) employing machine-learned interatomic potentials (MLIPs) serve as an efficient, urgently needed complement to \emph{ab initio} molecular dynamics (aiMD).
By training these potentials on data generated from \textit{ab initio} methods, their averaged predictions can exhibit comparable performance to \textit{ab initio} methods at a fraction of the cost.
However, insufficient training sets might lead to an improper description of the dynamics in strongly anharmonic materials, because critical effects might be overlooked in relevant cases, or only incorrectly captured, or hallucinated by the MLIP when they are not actually present.
In this work, we show that an active learning scheme that combines MD with MLIPs (MLIP-MD) and uncertainty estimates can avoid such problematic predictions.
In short, efficient MLIP-MD is used to explore configuration space quickly, whereby an acquisition function based on uncertainty estimates and on energetic viability is employed to maximize the value of the newly generated data and to focus on the most unfamiliar but reasonably accessible regions of phase space.
To verify our methodology, we screen over 112 materials and identify 10 examples experiencing the aforementioned problems. 
Using CuI and AgGaSe$_{2}$ as archetypes for these problematic materials, we discuss the physical implications for strongly anharmonic effects and demonstrate how the developed active learning scheme can address these issues.
\end{abstract}

\maketitle

\section{\label{sec:intro}introduction}
Machine-learned interatomic potentials (MLIPs) have an immense promise to accelerate molecular dynamics (MD) simulations since, in principle, they provide an accuracy nearing that of \textit{ab initio} calculations at a fraction of the cost~\cite{Lorenz:2004, Deringer:2019, Mishin:2021, Behler:2021}.
In material science, important applications for instance include thermal transport~\cite{Sosso:2012, Qian:2019, Korotaev:2019, Mortazavi:2020, Mangold:2020, Li:2020, Li:2020v2, Liu:2021, Verdi:2021, Langer:2023}, the dynamics of amorphous structures~\cite{Sosso:2012v2, Deringer:2017, Deringer:2018, Sivaraman:2020}, and ionic diffusion~\cite{Li:2017, Wang:2020, Shao:2021, Hajibabaei:2021}.
Recent works in this field have focused on improving these potentials by either decreasing their cost (faster running) or increasing their reliability for materials science applications (more reliable training).
Further accelerating the potentials for large-scale applications~\cite{Goryaeva:2019, Xie:2023} is necessary given that they are --albeit faster than first-principles approaches-- still considerably more costly than traditional force fields. 
For the latter case, this includes improvements in the MLIPs such as using the Euclidean group equivariant neural networks (e3nn)~\cite{Batzner:2022, Batatia:2022, Deng:2023, Yu:2024}, long-range interacting physics~\cite{Grisafi:2019, Zhang:2022, Gao:2022, Jaffrelot:2023, Anstine:2023}, or an enormously augmented amount of training data~\cite{Merchant:2023}. 

The key aspects of MLIP applications are the training process and the creation and selection of the data used for the training. 
Given that data production  --often performed via {\it ab initio} molecular dynamics (aiMD)-- can easily become computationally limiting, it is desirable to avoid redundancy,~i.e.,~the creation of additional data for areas that are already well covered in the training set.
Concurrently, it is pivotal, but impossible to guarantee, that the training data appropriately covers the relevant configurational space that will later be explored in the MD simulations.
This reflects the finding that MLIPs are most reliable in those areas for which enough training data is provided~\cite{Behler:2007, Bartok:2010, Wang:2018, Novikov:2021, Batzner:2022}.
In machine learning, this is phrased that the training data need to be independent and identically distributed (iid).
This is hard, if not impossible, to know.
Therefore, the active-learning scheme explained in the following paragraphs is critical.

To streamline training data production, several active-learning~($\mathcal{AL}$) approaches have been proposed in recent years.
These have been devised to address the aforementioned challenges associated with domain applicability, leading to a more effective and faster process of MLIP training~\cite{Mockus:1974, Dasgupta:2008}.
The key idea of $\mathcal{AL}$ is an iterative training of MLIP models by explicitly augmenting the training set with ``unfamiliar'' data, to achieve uniform reliability across the configurational space and avoid redundant learning for the well-trained areas.
Thus, the \emph{exploration} of the configurational space and \emph{sampling} of ``unfamiliar'' data are significant steps in the $\mathcal{AL}$ scheme.
To cope with the issue concerning exploration coverage, the MD simulation employing an efficient MLIP model (MLIP-MD) has been recently introduced to complement expensive aiMD~\cite{Zhang:2019, Zhang:2020, Carrete:2023, Xie:2021, Xie:2023v2, Zhu:2023, vanderOord:2023, Zaverkin:2023, Kulichenko:2023}.
It enables rapid exploration of vast spaces, leading to ample coverage of configurational space.
Furthermore, various methods for sampling these configurations as unfamiliar have been developed using metrics such as the similarity of the atomic environment~\cite{Li:2015}, a density-based hierarchical clustering~\cite{Sivaraman:2020}, and uncertainty estimates of MLIP models~\cite{Zhang:2019, Zhang:2020, Carrete:2023, Podryabinkin:2017, Verdi:2021, Xie:2021, Xie:2023v2, Zhu:2023}.
The combination of exploration and data-sampling methods allows for the retraining of MLIP models with unfamiliar data through on-the-fly~\cite{Li:2015, Sivaraman:2020} or iterative procedures~\cite{Zhang:2019, Zhang:2020, Carrete:2023, Xie:2021, Xie:2023v2, Zhu:2023, vanderOord:2023, Zaverkin:2023, Kulichenko:2023}.
In addition, novel exploration methods with uncertainty-biased dynamics have appeared in recent years to expedite the exploration of regions with high uncertainty~\cite{vanderOord:2023, Zaverkin:2023, Kulichenko:2023}.

The power of the described $\mathcal{AL}$ approaches in accelerating the training of more accurate MLIPs has been demonstrated for several applications.
To this end, it has been shown that such $\mathcal{AL}$ improved MLIPs and typically achieved a better description of microscopic quantities,~e.g.,~mean absolute errors~(MAE) of total energies, forces, stresses, etc.~\cite{Chen:2022, Morrow:2023, Chen:2023, Merchant:2023, Riebesell:2024} compared to {\it ab initio} reference data.
In turn, predictions of thermodynamic equilibrium properties,~e.g.,~temperature and pressure-dependent elastic constants, bulk moduli, phonon dispersions, radial-distribution functions, etc. also improve.~\cite{Li:2015, Sivaraman:2020, Zhang:2019, Zhang:2020, Carrete:2023, Xie:2021, Xie:2023, Zhu:2023, vanderOord:2023, Zaverkin:2023}.

In principle, $\mathcal{AL}$ is expected to improve the prediction of thermodynamic equilibrium and non-equilibrium properties.
For instance, the key aspect of transport coefficient calculations is how well MLIPs capture anharmonic lattice dynamics.
As described previously, MLIPs well describe the equilibrium lattice vibrations around the ground-state position, offering the long-term dynamics of the system and its memory.
From this, transport properties in most cases can be well determined by the time-autocorrelation of the respective fluxes in thermodynamic equilibrium as formulated in the fluctuation-dissipation theorem viz. the Green-Kubo formalism~\cite{Green:1954, Kubo:1957, Kubo:1957v2}.
However, in practice, materials undergo rare events that may disruptively impact phase transitions, local (phase) changes, and transport phenomena.
As examples of such non-equilibrium cases, it has been recently shown that the spontaneous formation of defects in CuI and of phase transition precursors in KCaF$_{3}$ are important dynamical phenomena that induce strong anharmonic behaviors in materials~\cite{Knoop:2023v1} resulting,~e.g.,~in a reduction of the thermal conductivity of CuI by a factor of 3.5.
An MLIP must be able to reproduce these anharmonic effects, but the ability of the $\mathcal{AL}$ addressing rare events is still elusive.

Naturally, one would assume that $\mathcal{AL}$ schemes would also improve predictions with respect to such rare events, as demonstrated,~e.g.,~for bond-breaking events in simple molecules.~\cite{Jung:2024}
However, a systematic quantification of the benefits of $\mathcal{AL}$ for complex materials has, so far, remained elusive.
In principle, a systematic comparison of {\it ab initio} and MLIP-MD calculations for transport calculations would be able to shed light on these questions.
However, this would require prohibitive numerical efforts for the first-principles MD to be able to reach the necessary statistics and time- and length scales relevant to strongly anharmonic events.
Similarly, one cannot explicitly target and monitor the phase space region associated with strongly anharmonic events since their occurrence and the associated path on the potential-energy surface are usually unknown {\it a priori}.
This is further aggravated by the fact that such events are typically short-lived, so that their influence on average properties like the MAE of microscopic quantities or predictions on thermodynamic equilibrium properties is small and hence hardly stands out against statistical noise. 

As demonstrated in this work, the problematic prediction of MLIPs regarding strongly anharmonic effects can happen in various contexts:
\begin{itemize}
    \item Rare events may be missing in the training data.
    \item Rare events may be present but with insufficient information about probabilities and lifetimes.
    \item Rare events may be present but smoothened away by regularization.
    \item MLIPs may well exhibit fake rare vents.
\end{itemize}
Alarmingly, such problems can be easily overlooked, since neither checking average predictions for testing data, such as mean absolute errors or $R^{2}$, nor inspecting close-to-equilibrium properties such as quasi-harmonic phonons allows to reliably detect these problems.

In this study, we build on existing $\mathcal{AL}$ ideas and adapt them for the description of strongly anharmonic materials by combining an ensemble uncertainty metric with thermodynamically meaningful acquisition functions.
Our benchmark on 112 materials for which extensive aiMD data is available from literature reveals that 10 out of these 112 materials require an $\mathcal{AL}$ approach to achieve a physically correct description of the PES.
We carefully analyze the failure of standard MLIP training for two representative examples.
In Sec.~\ref{sec:res-2},  we discuss CuI, the anharmonicity of which is severely underestimated procedures and in Sec.~\ref{sec:res-3} we discuss AgGaSe$_{2}$, the anharmonicity of which is strongly overestimated by standard MLIP training procedures.
Furthermore, we analyze how the proposed $\mathcal{AL}$ scheme rectifies these failures and propose best practices to achieve stable anharmonic MLIPs with $\mathcal{AL}$.
Eventually, Sec.~\ref{sec:physics} discusses the physical implication of the proposed approach for actual material-science predictions and shows that only the usage of $\mathcal{AL}$ guarantees to correctly identify the correct transport regime in actual MLIP-MD simulations.

\section{\label{sec:metho}Methodology}
In this section, we will first give a concise overview of the different established $\mathcal{AL}$ strategies employed in literature for the training of MLIPs in Sec.~\ref{sec:ALstr}.
In the following, Sec.~\ref{sec:ALappl} explains how we build on these concepts and adapt them to specifically target and benchmark strongly anharmonic materials.

\subsection{\label{sec:ALstr}State-of-the-Art $\mathcal{AL}$ strategies}

\subsubsection{\label{sec:stndALflow}The standard $\mathcal{AL}$ workflow}

\begin{figure}
\includegraphics[width=0.98\columnwidth]{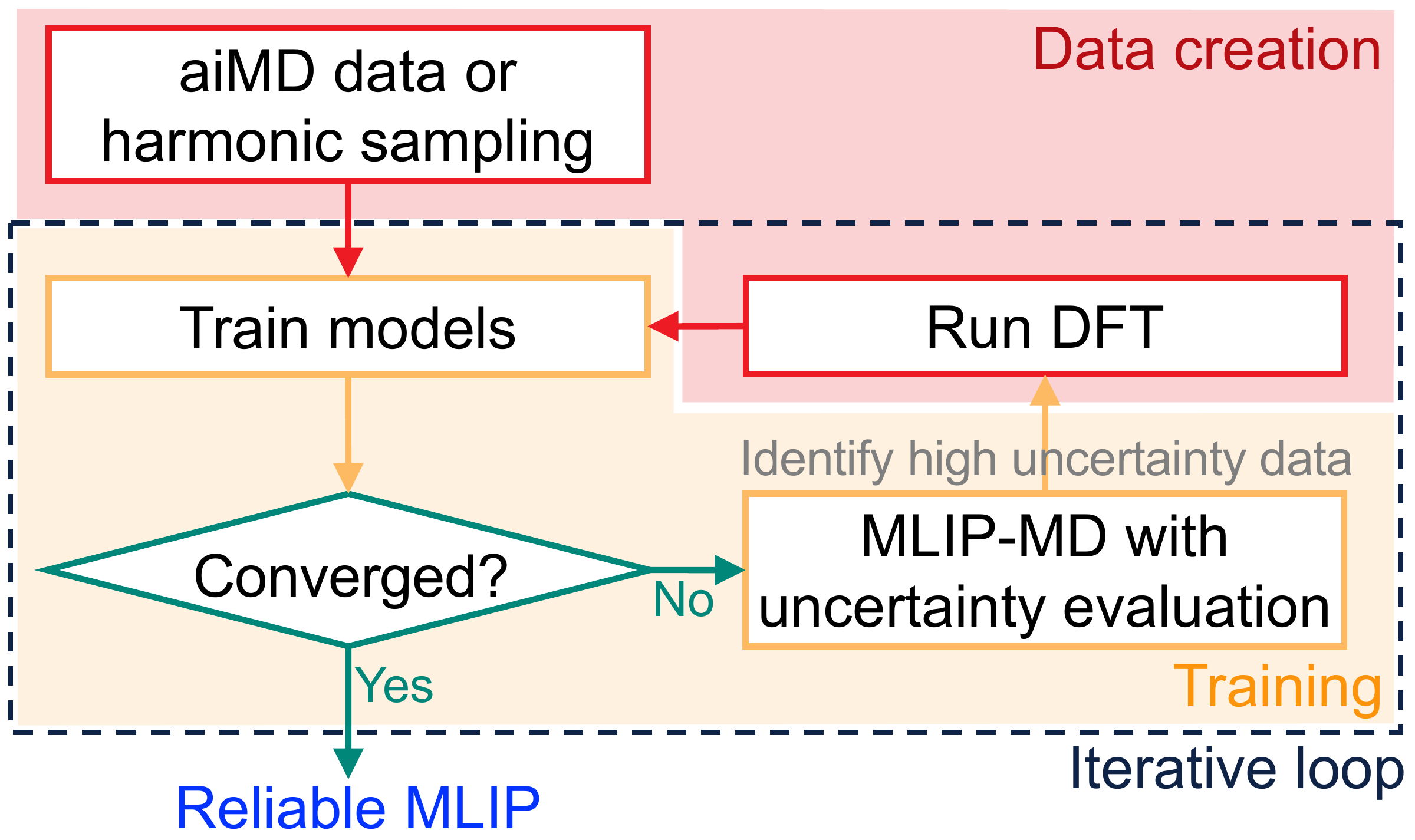}
\caption{\label{fig:workflow}
A workflow plot depicting the active learning ($\mathcal{AL}$) scheme.
}
\end{figure}

Here we summarize a step-by-step description of the five steps in a typical $\mathcal{AL}$ workflow, as illustrated in Fig.~\ref{fig:workflow}.
\begin{enumerate}
    \item {\bf Initialization}: An initial training set of configurations for a material is generated by either a short aiMD trajectory~\cite{Langer:2023}, stochastic sampling of phonon eigenvectors (harmonic sampling)~\cite{West:2006, Castellano:2022}, or random displacements of randomly chosen atoms~\cite{Takahasi:2017, Carrete:2023}.
    Because the $\mathcal{AL}$ process will improve their reliability anyway, even using pre-trained MLIPs, such as recent universal MLIPs~\cite{Batatia:2024, Deng:2023, Chen:2022} (See Fig.\,S1 in the supplementary material (SM)~\cite{Supplement} for more details), will work, unless their descriptions are incompatible with MD simulations.

    \item {\bf (Re)Training}: Initial(augmented) training data are utilized to (re)train MLIP models.

    \item {\bf Convergence test}: Once trained, the prediction quality of the MLIP is evaluated using a test set obtained from unseen parts of aiMD trajectories and error metrics such as the mean absolute errors (MAE).
    If the desired reliability is achieved, the iterative $\mathcal{AL}$ terminates, otherwise it proceeds to the next step.
    However, it is not able to ensure that a rare event pops up later.
    In addition, this reliability during the $\mathcal{AL}$ does not hold for temperatures higher than the trained temperature.

    \item {\bf Exploration and data-sampling}: The most critical step of the $\mathcal{AL}$ is the exploration of configurational space to find unfamiliar regions.
    Three popular methods to perform the exploration are aiMD~\cite{Li:2015, Sivaraman:2020}, explorative MLIP-MD~\cite{Zhang:2019, Zhang:2020, Carrete:2023, Xie:2021, Xie:2023v2, Zhu:2023}, or uncertainty-biased MD~\cite{vanderOord:2023, Zaverkin:2023}.
    For those approaches using MLIP, models trained with the data from previous iterations data are used in this step.
    Popular choices to sample each new configuration as familiar or not include the evaluation of extrapolation grades~\cite{Podryabinkin:2019}, the analysis of the structural similarity of samples~\cite{Li:2015}, the examination of correlations among samples~\cite{Sivaraman:2020}, and the uncertainty estimates of MLIP predictions~\cite{Zhang:2019, Zhang:2020, Carrete:2023, Xie:2021, Xie:2023v2, Zhu:2023, vanderOord:2023, Zaverkin:2023}.
    
    \item {\bf Data acquisition}: Full \textit{ab initio} calculations are performed on the snapshots sampled as unfamiliar to obtain their genuine force, energy, and stresses.
    This data is then added to the training set for the subsequent retraining.
    The workflow then goes back to the step 2.
\end{enumerate}
Steps 2-5 form a closed loop.
Whenever MLIP-MD, even during the practical applications, shows a high uncertainty estimate, the iterative $\mathcal{AL}$ scheme should resume.
The major difference among various $\mathcal{AL}$ schemes originates from the choice of exploration and data-sampling methods depending on their target systems.
For the exploration approaches, aiMD snapshots can be directly used as training data, whereas explorative MLIP-MD and uncertainty-based MD can effectively travel to unseen areas with their efficient implementations.
The choice of the exploration method is also influenced by the trade-off between cost and accuracy: aiMD trajectories are computationally expensive, but MLIP-MD may explore physically unfavorable parts of configurational space. 
The choice of data-sampling methods also depends on training environments and target properties, and more details are discussed in the following section.

\subsubsection{\label{sec:EnsmUncert}Ensemble uncertainty estimates}
\begin{figure}
\includegraphics[width=0.98\columnwidth]{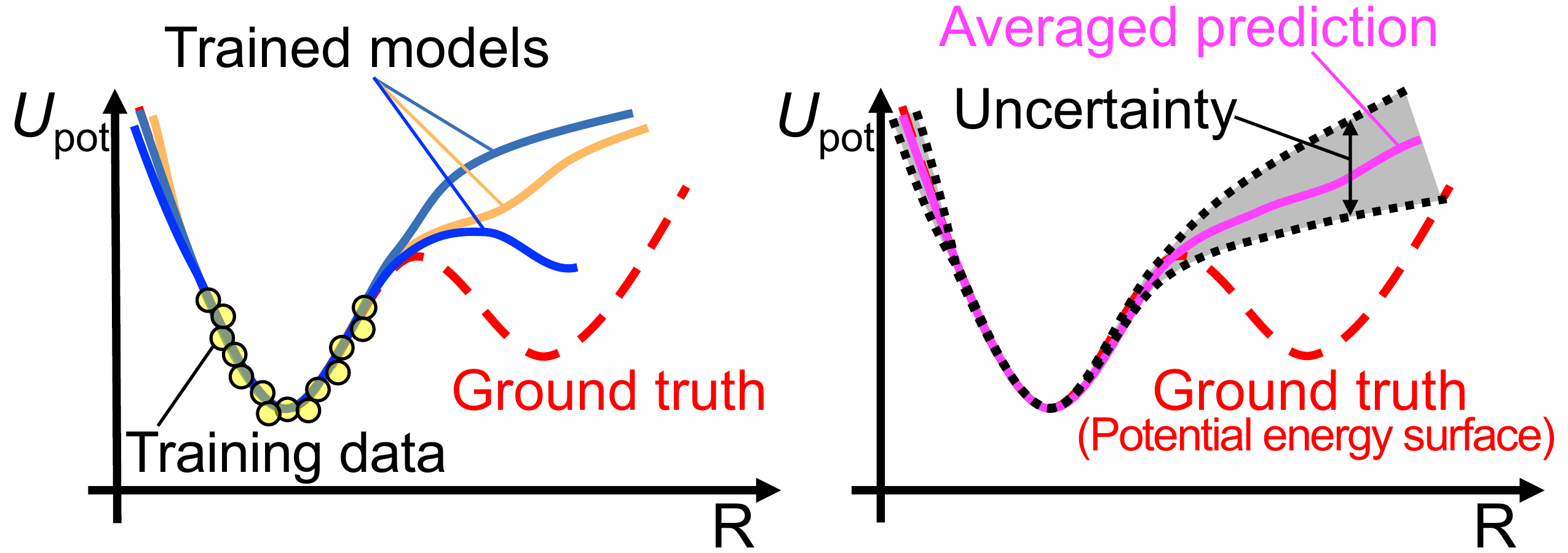}
\caption{\label{fig:al}
A schematic plot illustrating the definition of an ensemble uncertainty estimate in terms of the potential energy.
}
\end{figure}

Selecting an efficient and reliable \emph{data-sampling} method is critical to the performance of the $\mathcal{AL}$ scheme as that determines how effectively it identifies unfamiliar data among the substantial volume of new data generated.
In this study, the uncertainty estimates of the MLIP predictions serve as a qualitative signal for when the MD trajectory leaves the well-trained regions.
Although concerns persist regarding its quantitative usage in extrapolation regions~\cite{Kahle:2022}, uncertainty estimates still hold remarkable value as a qualitative indicator.
Various uncertainty techniques have been developed such as uncertainty estimates from the probabilistic framework of Bayesian linear regression~\cite{Podryabinkin:2017, Verdi:2021} or an internal principled uncertainty quantification mechanism to evaluate the prediction uncertainty~\cite{Xie:2021, Xie:2023v2}.
These approaches can quickly assess the uncertainty estimates and do not require training multiple models.
However, this approach can only be applicable to the MLIP models, which can internally provide a probabilistic model, e.g., the Gaussian mixture model~\cite{Zhu:2023}.

Instead, this section introduces an ensemble uncertainty estimate in detail due to its simple implementation and wide applicability across various MLIP architectures.
The ensemble uncertainty estimate is assessed as a standard deviation of predictions from multiple MLIP models that are trained using different training sets (subsampling) and different initial structures and parameters of the neural networks (deep ensemble), as depicted in Fig.~\ref{fig:al}.
The resulting MLIP models consistently predict reliable atomic motions within well-trained areas but begin to diverge beyond these areas, as illustrated in Fig.~\ref{fig:al}.
Despite the fact that such uncertainty estimates do not allow for a quantitatively exact prediction of the error and, hence, of the ground-truth potential-energy surface~\cite{Kahle:2022, Mendes:2024,Lahlou:2023},
higher than average uncertainty estimates can be used to qualitatively identify unfamiliar regions so to steer further sampling for retraining MLIP models.

Here we describe how to obtain ensemble uncertainties of the potential energy and forces employed as the data-sampling method in the $\mathcal{AL}$ scheme.
The uncertainty estimate for a target property ($X$) of a MLIP-MD snapshot is determined by calculating the standard deviation of predicted target properties using the following equation:
\begin{equation}
\label{eq:Uncert}
    \mathrm{UCE}_{X} = \sqrt{\frac{1}{N}\sum^{N}_{I} \left ( X_{I} - \mu_{X} \right )^{2}}.
\end{equation}
Here, $X_{I}$ represents the target energy predicted by the $I^{\mathrm{th}}$ MLIP model, and $\mu_{X}$ corresponds to the mean value of target properties from the ensemble of all $N$ different MLIP models.
Thus, the potential-energy uncertainty estimate~($\mathrm{UCE}_{U}$) is obtained from the standard deviation with respect to a predicted mean value of potential energies ($U$).
The maximum of the predicted forces uncertainty estimates $\left ( \mathrm{UCE}^{\mathrm{max}}_{F} \right )$ is evaluated as the largest one among the uncertainties estimate $\left ( \mathrm{UCE}^{i}_{F} \right )$ of each force for all different $i^{\mathrm{th}}$ atoms in the material ($\mathbf{F}^{i}_{I}$).

\subsection{\label{sec:ALappl}Adaption of $\mathcal{AL}$ to strongly anharmonic systems}

\subsubsection{\label{sec:ExplLab}Exploration and data-sampling methods}
Our $\mathcal{AL}$ scheme is designed to iteratively retrain MLIP models with unfamiliar strongly anharmonic events selectively sampled during configurational space exploration.
Since such events can occur infrequently during MD explorations, aiMD might not be able to reach the relevant time and length scales to sample them properly or capture them at all.
Instead, explorative MLIP-MD is used as an efficient complement to aiMD, enabling simulations to reach the needed time and length scales to observe also rare events and estimate the model's uncertainty.
In this study, we implement MLIP-MD based on the mean value of the forces from MLIP models, resulting in one MLIP-MD trajectory with target uncertainty estimates.

Ensemble uncertainty estimates are selected as the data-sampling method in the present study due to its generalized application regardless of MLIP types.
Ensemble uncertainties of total energy and maximum atomic force in Sec.\,\ref{sec:EnsmUncert} are examined to verify its ability to sense the unfamiliar strongly anharmonic events.
We also extend this approach to the degree of anharmonicity because this quantity can effectively identify the anharmonic rare event using a single value.
The degree of anharmonicity ($A$) at the MD simulation time ($t$) is defined based on a concept devised by Knoop \emph{et al.} as follows~\cite{Knoop:2020}.
\begin{equation}
\label{eq:anhar}
A(t) = \sqrt{\frac{\sum_{i,\alpha} \left ( F^{i,\alpha}_{\mathrm{Aha}}(t) \right )^{2} }{\sum_{i,\alpha} \left ( F^{i,\alpha}(t) \right )^{2} }},
\end{equation}
where $F^{i,\alpha}(t)$ represents the $\alpha$~$(= x,y,z)$ component of the force vector on the $i$-th atom at the MD time ($t$).
$F^{i,\alpha}_{\mathrm{Aha}}(t)$ means the $\alpha$ component of the anharmonic atomic force vector on the $i$-th atom at the time ($t$), evaluated as $F^{i,\alpha}_{\mathrm{Aha}}(t) = F^{i,\alpha}(t) - F^{i,\alpha}_{\mathrm{Ha}}(t)$, where $F^{i,\alpha}_{\mathrm{Ha}}(t)$ represents the $\alpha$ component of the force that would be obtained at the same geometry but using a harmonic (parabolic) potential.
Accordingly, the uncertainty of the degree of anharmonicity of a configuration $\left ( \mathrm{UCE}_{A} \right )$ is defined as the standard deviation with respect to the degree of anharmonicity ($A$) via Eq.\,\ref{eq:Uncert}.

\subsubsection{\label{sec:SampProb}Sampling probability}
\begin{figure}
\includegraphics[width=0.98\columnwidth]{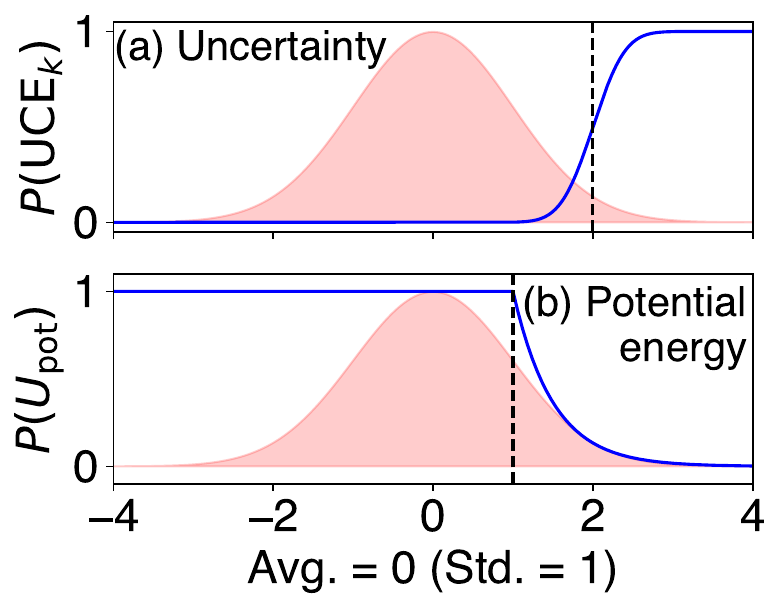}
\caption{\label{fig:prob}(Color online.)
Sampling probability distribution (blue solid lines) regarding (a) uncertainty estimates and (b) potential energy in relation to a reference average set at 0, with each integer representing a standard deviation of 1.
The pink bell-shaped plots represent schematic normal distributions.
}
\end{figure}

For retraining, it is necessary to balance between maximizing the uncertainty estimate of the targeted regions of phase-space and ensuring those regions are at least accessible. 
To this end, we propose a sampling probability that serves as an acquisition function.

First, data-sampling new training data for subsequent rounds of MLIP training necessitates the definition of high uncertainty estimates.
The uncertainty estimate fluctuates across different $\mathcal{AL}$ steps due to the changes in training data and training areas.
Hence, high uncertainty estimates in our $\mathcal{AL}$ scheme mean a \emph{relatively higher} uncertainty estimate compared to other data points.
At the beginning of each $\mathcal{AL}$ step, we evaluate uncertainties in 300 testing data randomly selected from aiMD trajectories with 12001 snapshots (not included in the training set), recording their mean value and standard deviation.
These quantities serve as reference points to identify unfamiliar MD snapshots from preliminary MLIP-MD.
Then, we set a soft selection criterion by building a sampling probability function in terms of uncertainty estimates, shown as $P(\mathrm{UCE}_{k})$, creating a smooth curve, a cumulative normal distribution function, at the criterion limit at two standard deviations away from the average (dashed line in Fig.\,\ref{fig:prob} (a)).
The mathematical form of this criterion, $P(\mathrm{UCE}_{k})$, is described below:
\begin{equation}
\label{eq:critria_Uncert}
P(\mathrm{UCE}_{k}) = \frac{1}{2} \left [ 1 + \mathrm{erf} \left ( \frac{\mathrm{UCE}_{k} - \mu_{\mathrm{UCE}^{\mathrm{test}}_{k}} - 2\sigma_{\mathrm{UCE}^{\mathrm{test}}_{k}}}{\sigma_{\mathrm{UCE}^{\mathrm{test}}_{k}} \sqrt{0.2}} \right ) \right ].
\end{equation}
Here, $\mathrm{UCE}_{k}$~$(=\mathrm{UCE}_{U}, \mathrm{UCE}^{\mathrm{max}}_{F},$ or $\mathrm{UCE}_{A})$ signifies the uncertainty estimate at the current step.
$\mu_{\mathrm{UCE}^{\mathrm{test}}_{k}}$ and $\sigma_{\mathrm{UCE}^{\mathrm{test}}_{k}}$ mean the average and standard deviation of uncertainties in testing data, respectively.
Fig.~\ref{fig:prob} (a) illustrates that $\mathcal{AL}$ mostly collects data with uncertainty estimates larger than two times the standard deviation away from the reference average of testing data.
The probability function shape is adopted from a cumulative normal distribution function, yielding a soft limit rather than a rigid threshold.

Second, we considered the potential energy of sampled data to prevent sampling unphysical configurations that have enormously high energy.
This criterion is implemented by a different probability function, similar to the one for uncertainty estimates.
The sampling probability function regarding the potential energy, $P(U)$, is derived from a modified probability distribution of the canonical ensemble by decaying the probability from a criterion limit at one standard deviation away from the average (dashed line in Fig.\,\ref{fig:prob} (b)).
In the mathematical expression,
\begin{equation}
P(U) =
\begin{cases}
1 \quad\quad\quad \text{if } U \leq \mu_{{U}^{\mathrm{test}}} + \sigma_{U^{\mathrm{test}}},\\
\exp\left( \frac{\ln{0.2}}{0.8} \cdot \frac{U - \mu_{U^{\mathrm{test}}} - \sigma_{U^{\mathrm{test}}}}{\sigma_{U^{\mathrm{test}}}} \right) \\
\quad\quad\quad\;\: \text{if } U > \mu_{{U}^{\mathrm{test}}} + \sigma_{U^{\mathrm{test}}} ,
\end{cases}
\end{equation}
where $U$ represents the potential energy at the current step of the MD simulation.
$\mu_{{U}^{\mathrm{test}}}$ and $\sigma_{U^{\mathrm{test}}}$ mean the average and standard deviation of potential energies predicted by MLIP models in testing data.
The coefficients are determined to ensure that $P(U)$ starts decaying from the criterion limit,~i.e.,~at one standard deviation away from the average.
Fig.~\ref{fig:prob} (b) displays that this criterion starts to exclude data with potential energy beyond one standard deviation away from the reference average.
Finally, the actual sampling process is executed based on the combined criteria of both probability functions, concluding $P = P(\mathrm{UCE}_{k}) \cdot P(U)$.

Accordingly, the proposed $\mathcal{AL}$ scheme augments state-of-the-art $\mathcal{AL}$ workflows with a soft sampling criterion that ensures that strongly anharmonic effects are correctly captured, even if they occur infrequently.

\section{\label{sec:comp}Computational details}

For comprehensive analysis and comparison, we applied our $\mathcal{AL}$ workflow to 112 different bulk materials from the NOMAD aiMD repository, which contains extended MD simulations up to 60~ps for thermal transport studies previously conducted by Knoop \emph{et al.}~\cite{Knoop:2023v1, Knoop:2023data}.
Since the consistent training of MLIP models should be ensured to attain their reliable predictions, the sampled data during the $\mathcal{AL}$ iterations are calculated employing exactly the same computational frameworks.
In detail, all DFT calculations were executed by the \texttt{FHI-aims} code packages employing an all-electron formalism~\cite{Blum:2009, Knuth:2015}.
The utilized supercells with 160-256 atoms are consistently chosen to describe the dynamics of 112 materials during the $\mathcal{AL}$ workflow.
The identical $\mathbf{k}$-point sampling densities were applied to integrate the Brillouin zone.
The basis sets are set to \emph{light} default in the \texttt{FHI-aims}, and PBEsol~\cite{Perdew:2008} is selected as exchange-correlation functional.
All quasi-harmonic and anharmonic vibrational properties of of CuI and AgGaSe$_{2}$ are computed using \texttt{FHI-vibes}~\cite{Knoop:2020v2}; whereby the perturbative formalism implemented in \texttt{phonopy}~\cite{Togo:2023v1, Togo:2023v2} and \texttt{phono3py}~\cite{Togo:2015, Togo:2023v1} is used for the harmonic phonon dispersions and phonon lifetimes.
For more details, we refer to the original aiMD paper~\cite{Knoop:2023v1} and its data repository~\cite{Knoop:2023data}.

For the MLIP architecture, we adopt a recent graph neural network potential with a message-passing scheme implemented by NequIP version 0.5.6~\cite{Batzner:2022}.
Throughout convergence tests for 300 training data, the values of hyperparameters are carefully determined through the convergence test for potential energy and forces predictions and they are consistently applied for all $\mathcal{AL}$ benchmarks.
Our hyperparameters are a local cutoff radius for the atomic environment~(\texttt{rmax}) of 5~$\mathrm{\AA}$, a maximum rotation order for the neural network features~(\texttt{lmax}) of 3, and a feature multiplicity of 32, respectively.
Four layers of the neural network are employed, and eight basis functions are used in the radial basis.
To balance the prediction accuracy, the loss function ratio between potential energy per atom~(\texttt{PerAtomMSELoss}) and atomic forces is chosen to be 1:1.
The \texttt{float64} precision is adopted to maintain the high accuracy.
All MLIP training with NequIP is implemented via a single GPU core from the NVIDIA Tesla A100.
A total of six MLIP models are utilized to evaluate the ensemble uncertainty by training three different MLIP models using \emph{subsampled} training datasets with two distinct random initializations with a \emph{deep ensemble} approach, which provides the optimal performance by balancing more converged uncertainty estimates and the computational costs for running multiple MLIP models.
In the initialization, each MLIP model is trained with 25 training data and 5 validation data from the aiMD trajectory.
Sequentially, the DFT results of 30 new MLIP-MD data from the exploration and data-sampling step are added to the previous training data with a consistent ratio of training and validation.

Our iterative $\mathcal{AL}$ workflow is automatically implemented by a Python code, \texttt{ALmoMD}~\cite{almomd:2024}, that interfaces a DFT code (\texttt{FHI-aims}~\cite{Blum:2009, Knuth:2015}) and a MLIP code (\texttt{NequIP}~\cite{Batzner:2022}) via the Atomic Simulation Environment (\texttt{ASE})~\cite{Larsen:2017} and the \texttt{FHI-vibes}~\cite{Knoop:2020v2}.
In this study, the $\mathcal{AL}$ workflow is conducted based on the initial settings for NequIP and MLIP-MD as described above.
In practical applications, human intervention may be possible to tune the loss function ratio of NequIP or adjust the sampling criteria.
However, this study keeps the initial setting without any intervention during the $\mathcal{AL}$.
The MLIP-MD in the $\mathcal{AL}$ is implemented for the NVT ensemble employing the Langevin thermostat to explore the configurational space with a target temperature of 300\,K using the \texttt{ASE} library~\cite{Larsen:2017}.
The friction parameter and timestep of MLIP-MD are set to 0.03 and 5\,fs, respectively, to follow the original MD setting in the referenced repository~\cite{Knoop:2023data}.
The phonon lifetimes from the MD trajectory are extracted by the workflow designed in the \texttt{FHI-vibes}~\cite{Knoop:2020v2}.
Its MD simulation is implemented based on the NVE ensemble, and the detailed MD parameters are set the same as the one implemented in the aiMD data repository~\cite{Knoop:2023v1, Knoop:2023data}.
Nudged elastic band (NEB) calculations~\cite{Jonsson:1998, Corrales:1998, Henkelman:2000} are performed with the implementation available within \texttt{ASE}~\cite{Larsen:2017} to extract the potential energy surface between the ground state structure and the structure with a defect.

\section{\label{sec:res}Results and Discussion}

\begin{figure}
\includegraphics[width=0.98\columnwidth]{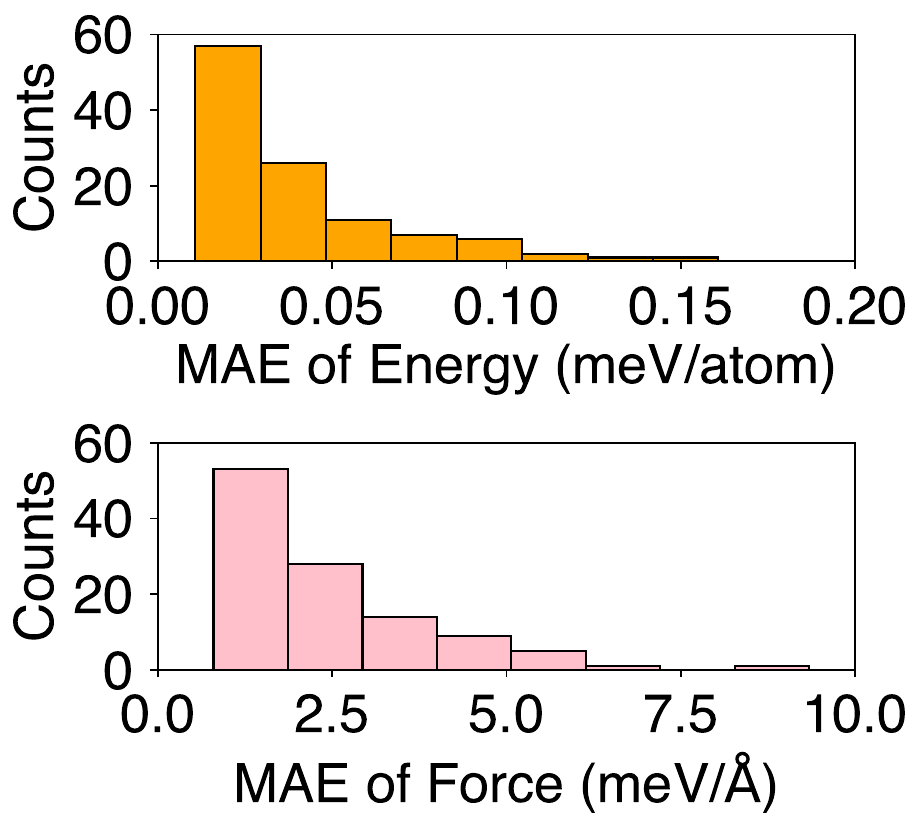}
\caption{\label{fig:data_MAE}
Mean absolute errors (MAE) statistics of energy (top) and atomic force (bottom) predicted by MLIP models from standard training metrics for 112 bulk materials.
The bin width is determined based on Sturges's rule~\cite{Sturges:1926}.
}
\end{figure}

To substantiate our argument regarding erroneous MLIP predictions associated with rare events and test our concept and the proposed $\mathcal{AL}$ approach, we first investigated 112 bulk materials with MLIP training via random sampling for the aiMD trajectory,~i.e.,~without $\mathcal{AL}$. 
Fig.\,\ref{fig:data_MAE} shows the mean absolute error results of energy and atomic force predictions from MLIP models using the ordinary training approach.
All results for 112 bulk materials show high-accuracy predictions for the energy ($<$ 1\,meV/atom) and atomic forces ($<$ 10$^{-2}$\,eV/$\mathrm{\AA}$) of testing data, 
which was not seen during training. At first, this erroneously suggests that the trained MLIP models will provide reliable predictions for all these materials.

\begin{figure}
\includegraphics[width=0.98\columnwidth]{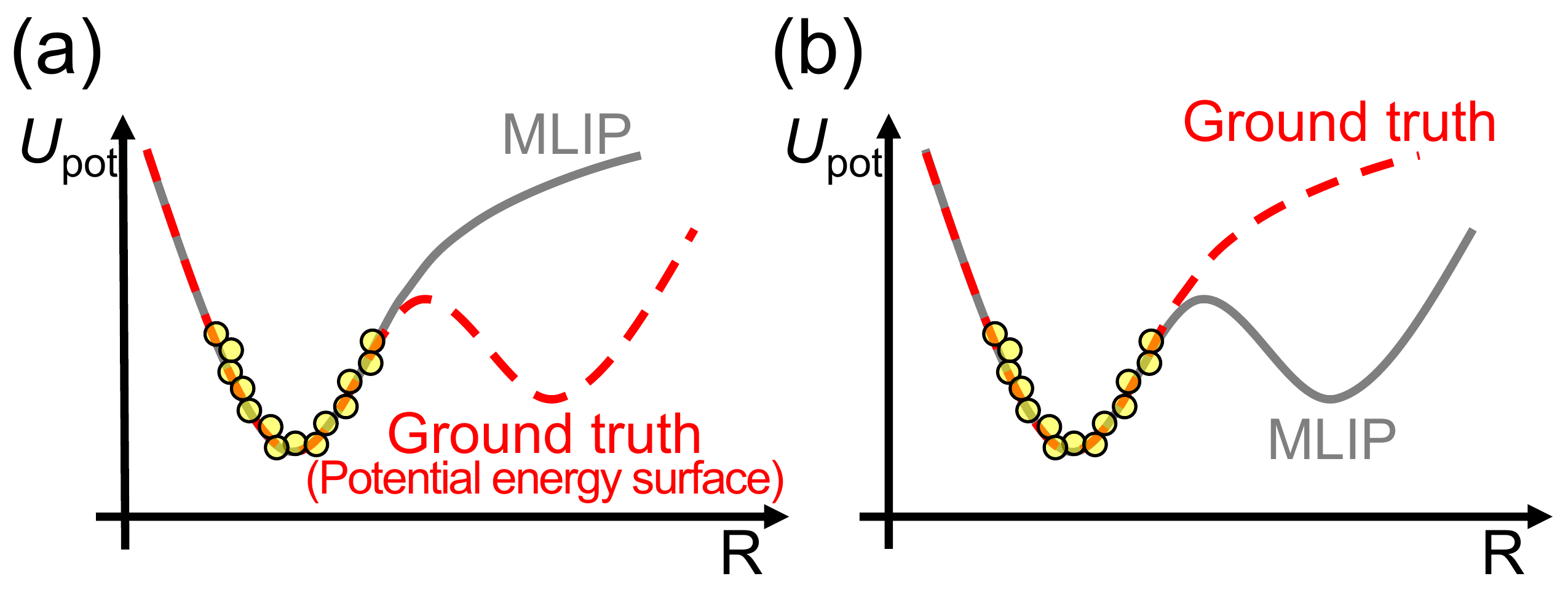}
\caption{\label{fig:case}
Schematic plots representing problematic scenarios: (a) the absence of metastable states in MLIP and (b) the prediction of erroneous metastable states in MLIP. The yellow circles represent the training data of MLIP models.
}
\end{figure}

However, at the considered temperature, 10 of these materials~($\approx 9$\%) required multiple $\mathcal{AL}$ iterations before reaching convergence, despite the fact that the MAE of these material were comparable 
to all other materials in the initial training.
A deeper analysis revealed that the MLIPs for these 10 materials featured erroneous predictions before AL in extrapolated regions not covered by the training set, as illustrated in Fig.\,\ref{fig:case}.
Here, we can distinguish between two distinct cases: The MLIP model may either overlook the presence of a metastable state (Fig.\,\ref{fig:case} (a)) or predict a false metastable state (Fig.\,\ref{fig:case} (b)) that is not present in actual first-principles calculations.
Both scenarios are associated with \emph{exploration} and \emph{sampling} during the standard training method.
A low visiting frequency for high-energy training boundary areas yields a low sampling of configurations in these regions.
Obviously, this could in principle be mitigated running extensively long aiMD simulations, which come with impractically increased computational effort.
But even in this case, human inspection would be needed to ensure that the respective phase-space regions associated with the rare event are appropriately covered in test and training sets.
In the following, we discuss the two scenarios using representative examples and show that the proposed $\mathcal{AL}$ is able to correct these erroneous predictions in an automatic fashion 
with modest computational overhead. Furthermore, materials that do not suffer under the described problems are not affected by the $\mathcal{AL}$ approach, as exemplarily shown for 
KCaF$_{3}$ in the SM~\cite{Supplement} (See Fig.\,S2 and S3).

\subsection{\label{sec:res-2} CuI: The Case of Missing Minima}
\begin{figure}
\includegraphics[width=0.98\columnwidth]{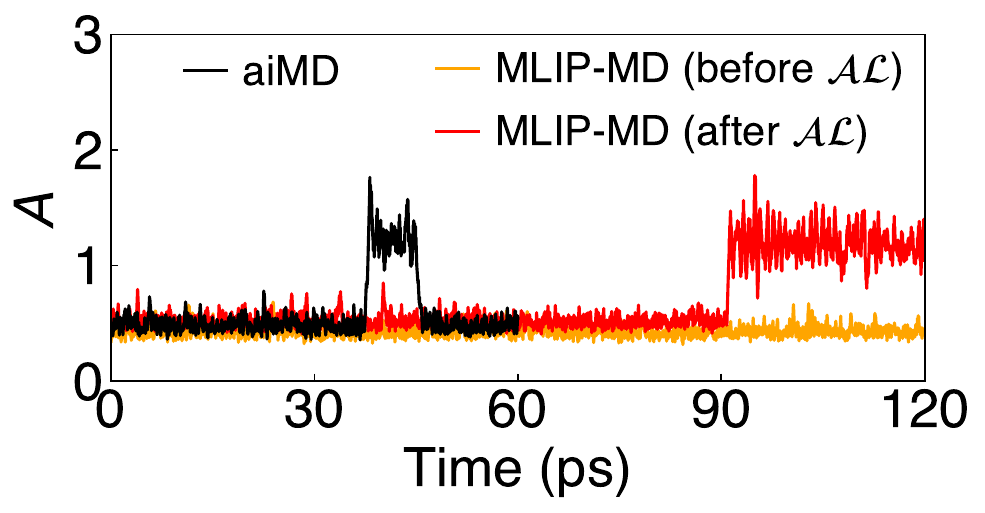}
\caption{\label{fig:mdtraj}
The degree of anharmonicity ($A$) of MD trajectories of CuI from aiMD (Ref.~\cite{Knoop:2023v1}) and MD using MLIPs before/after applying the $\mathcal{AL}$ scheme.
}
\end{figure}

For the first scenario, in which a metastable state associated with a strongly anharmonic effect is overlooked by the MLIP, we discuss the case of copper iodide~(CuI).
Already in a previous aiMD study, it was reported that CuI features spontaneous defect creation at 300\,K, which has a significant impact on the resulting thermal conductivity~\cite{Knoop:2023v1}.
However, defect creation occurs infrequently on aiMD time-scales,~e.g., it is observed --when at all-- only after more than 35~ps in Ref.~\cite{Knoop:2023v1} as illustrated as a black solid line in Fig.\,\ref{fig:mdtraj} and can happen at different any point for other trajectories.
When the degree of anharmonicity jumps from 0.5 to 1.2, it implies the occurrence of the defect creation.
This defect creation causes a strongly anharmonic effect, which impacts the transport properties; e.g., the phonon lifetime is drastically reduced by this effect.
Accordingly, an MLIP training procedure that only focuses,~e.g.,~on then first 10~ps of the trajectory or fitting by regularization, is prone to miss this mechanism, as discussed below.
Qualitatively, this strongly anharmonic effect leads to a breakdown of the phonon picture~\cite{Simoncelli:2022, Caldarelli:2022} as quantified by the Ioffe-Regel limit~\cite{Ioffe:1960}.
When a phonon lifetime becomes shorter than the oscillation period, the vibrational quasi-particle becomes invalid, as for instance observed for 
in the case of spontaneous defect creation~\cite{Knoop:2023v1}. Conversely, the quasi-particle picture holds when the lifetimes are longer and 
transport theories based on the phonon picture can be used to evaluate heat transport. Therefore, it is important to verify whether MLIP can reproduce such physics associated with strong anharmonicity.

\begin{figure}
\includegraphics[width=0.92\columnwidth]{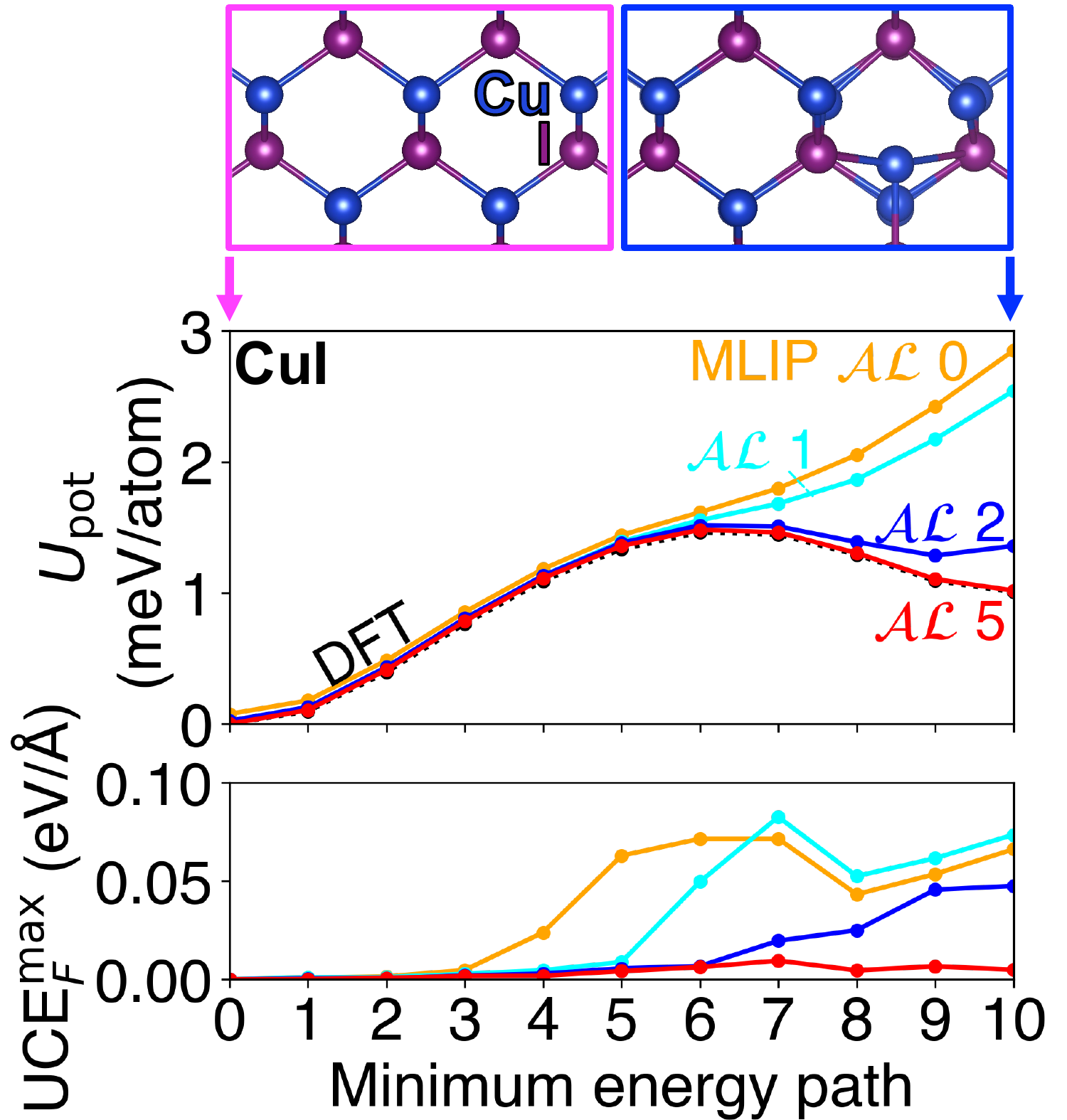}
\caption{\label{fig:CuI}
The PES and corresponding maximum uncertainty estimates of atomic forces ($\mathrm{UCE}^{\mathrm{max}}_{F}$) predicted by MLIP from $\mathcal{AL}$ using $\mathrm{UCE}^{\mathrm{max}}_{F}$ as a function of the gradual structural evolution of CuI, transitioning from its ground state structure (top left outset structure) to a defect-bearing configuration (top right outset structure) via NEB calculations. The black line shows the DFT ground truth and the colored points with solid lines represent PES results predicted by MLIP models. The $\mathcal{AL}$ $N$ means the PES snapshots from MLIP models obtained after the $N$-th step of the $\mathcal{AL}$ iterations. 
}
\end{figure}

Although, in the present case, we {\it a priori} know about the existence of the metastable defect state, we do not exploit this information for the initial training of the MLIP, so to mimic the typical application case
in which little is known about the different processes that might be active in a material at the beginning. Accordingly, we train the initial MLIP on one of the aiMD trajectories, which does not contain any explicit defect formation.
Accordingly, the resulting MLIP fails to predict a metastable state, as checked by running 30 independent MLIP-MD before $\mathcal{AL}$ for 1\,ns.
Energetically, this is rationalized in Fig.~\ref{fig:CuI}, in which the minimum-energy path between a pristine structure and a metastable defect is plotted both for the ground-truth DFT data and the different iterations~$N$ of the MLIP training labeled with $\mathcal{AL}$ $N$.
While the initial MLIP model ($\mathcal{AL}$ 0) erroneously misses the presence of a metastable state, subsequent $\mathcal{AL}$ iterations incorporate information about the respective phase-space region in the training data, so to correctly reproduce the DFT PES at the fifth iteration, at which convergence is achieved.
This is further substantiated by the fact that also the respective uncertainty estimates in the maximum forces shows a well-balanced behavior over the complete minimum energy path, cf.~the bottom panel in Fig~\ref{fig:CuI}.
In the MD trajectory of CuI, the defect creation missed in MLIP-MD before $\mathcal{AL}$ is now observed in MLIP-MD after $\mathcal{AL}$, as illustrated in Fig.\,\ref{fig:mdtraj}.
Let us emphasize that no information about the presence of the defect state was fed to the $\mathcal{AL}$ procedure; the observed improvements solely result automatically from the designed acquisition function that iteratively enables sampling high uncertainty regions, even if they appear to be energetically inaccessible at first, see Fig.~S3.

\begin{figure}
\includegraphics[width=0.875\columnwidth]{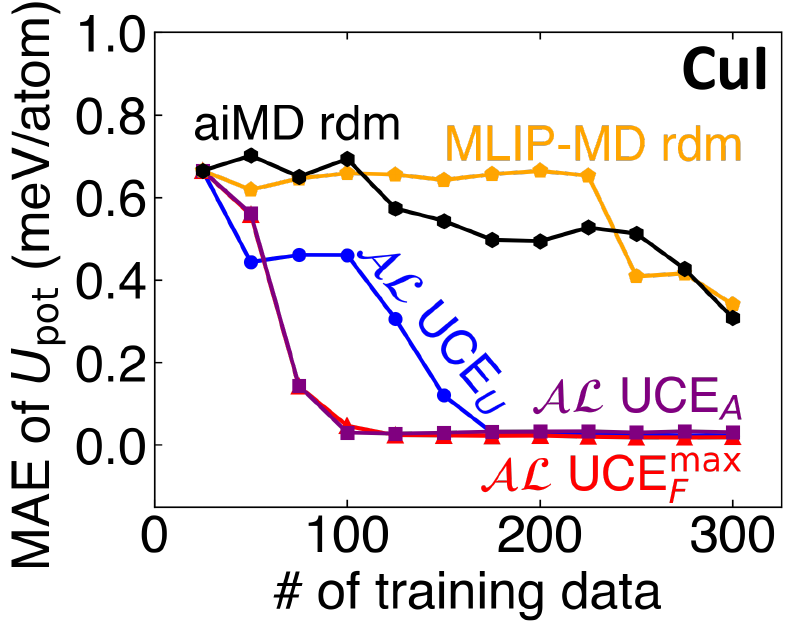}
\caption{\label{fig:CuI_MAE}
MAE of potential energies in testing data from aiMD trajectory of CuI as a function of the number of training data.
The testing data also includes configurations with defects that only occur rarely in the dynamics.
\texttt{aiMD rdm}~(black hexagon) involves MLIP training with random sampling from aiMD trajectories.
\texttt{MLIP-MD rdm}~(orange pentagon) represents MLIP training with random sampling from MD trajectories using MLIP at each step.
$\mathcal{AL}$~$\mathrm{UCE}_{U}$~(blue circle), $\mathcal{AL}$~$\mathrm{UCE}^{\mathrm{max}}_{F}$~(red triangle), and $\mathcal{AL}$~$\mathrm{UCE}_{A}$~(purple square) exhibit MLIP training with the active learning approach utilizing energy uncertainty estimates, maximum uncertainty estimates of atomic forces, and uncertainty estimates in the degree of anharmonicity, respectively.
}
\end{figure}

The acceleration of MLIP training via the $\mathcal{AL}$ scheme is also observed from the mean absolute error (MAE) check of testing results, as shown in Fig.\,\ref{fig:CuI_MAE}.
Fig.\,\ref{fig:CuI_MAE} depicts how the MAE of potential energies in testing data varies when we augment training data in five different methods.
\texttt{aiMD rdm} stands for random sampling of training data from aiMD trajectory, while \texttt{MLIP-MD rdm} follows the $\mathcal{AL}$ workflow using random sampling instead of data-sampling via uncertainty estimates.
The actual $\mathcal{AL}$ implementations utilize three different uncertainties ($\mathrm{UCE}_{U}$, $\mathrm{UCE}_{F}^{\mathrm{max}}$, and $\mathrm{UCE}_{\mathrm{A}}$) introduced in Sec.\,\ref{sec:EnsmUncert} and Sec.\,\ref{sec:ExplLab}.
Testing data comprise MD snapshots from aiMD trajectories, including strongly anharmonic events.
$\mathcal{AL}$ $\mathrm{UCE}_{F}^{\mathrm{max}}$ and $\mathcal{AL}$ $\mathrm{UCE}_{A}$ reach the convergence first, followed by $\mathcal{AL}$ $\mathrm{UCE}_{U}$.
This difference stems from the fact that $\mathrm{UCE}_{F}^{\mathrm{max}}$ has an atomic resolution of uncertainty evaluation and the degree of anharmonicity ($A$) used in $\mathrm{UCE}_{A}$ is sensitive to defect creation whereas $\mathrm{UCE}_{U}$ utilizes the potential energy into that all atomic information are merged.
\texttt{aiMD rdm} and \texttt{MLIP-MD rdm} trained with up to 300 training data could not get similar reliability and convergence behavior for energy predictions because they could not properly sample rare events from their trajectory.
From this, we could draw the lesson that implementing $\mathcal{AL}$ leads to effective MLIP training, even further effective when the data-sampling method has atomic resolution or structural change sensitivity.

\subsection{\label{sec:res-3}AgGaSe$_{2}$: The Case of Fictitious Minima}
\begin{figure}
\includegraphics[width=0.90\columnwidth]{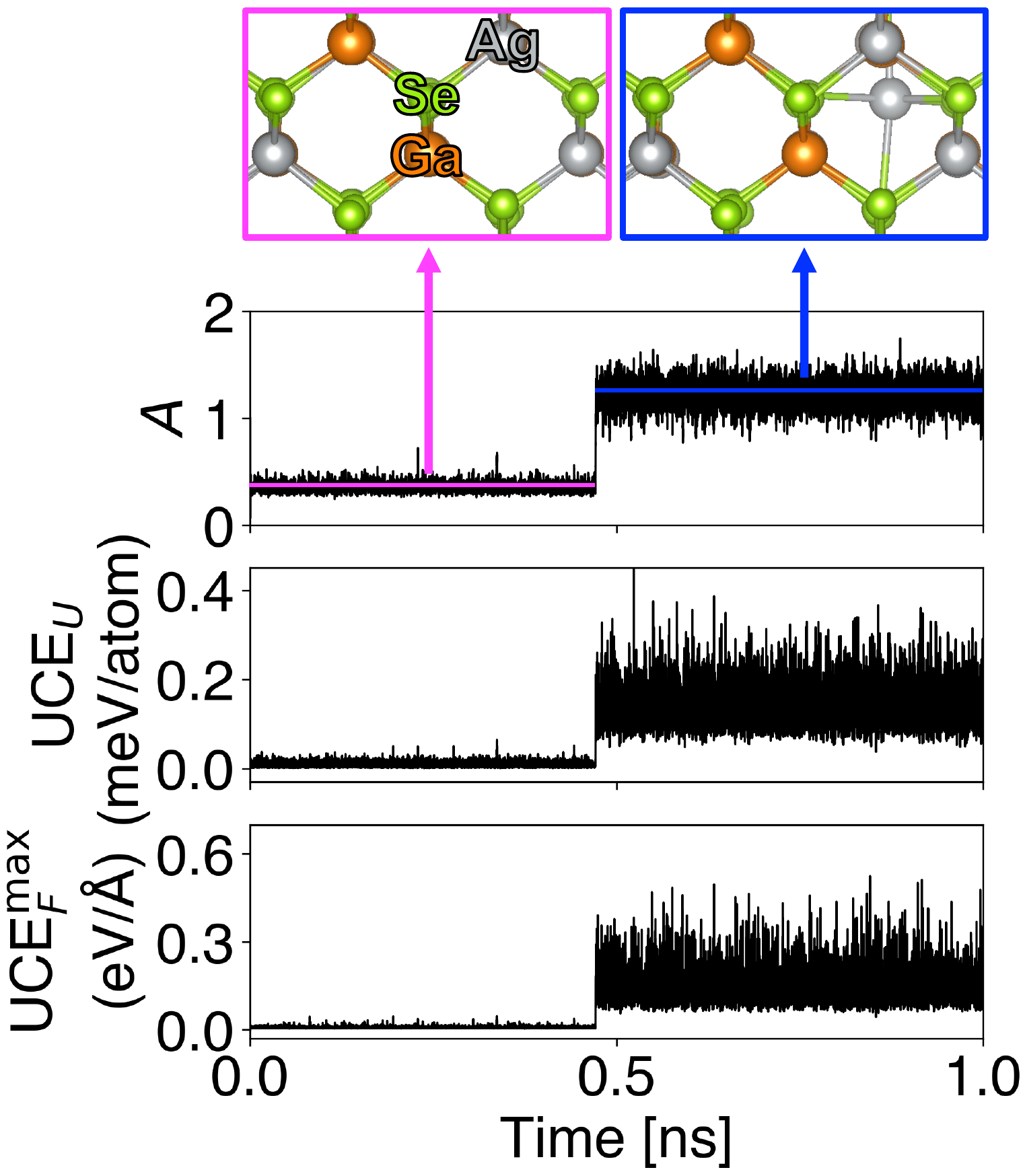}
\caption{\label{fig:AgGaSe2_aiMD}
The MD trajectory using MLIP trained on 300 aiMD data, showing the evolution of anharmonicity degree ($A$), energy uncertainty estimates ($\mathrm{UCE}_{U}$), and maximum uncertainty estimates of atomic force ($\mathrm{UCE}_{F}^{\mathrm{max}}$) over a 1~ns simulation.
}
\end{figure}

As a second scenario, we discuss the case of AgGaSe$_{2}$, for which the initial MLIP incorrectly predicts a metastable state that induces fictitious, strongly anharmonic effects that are not active on the ground-truth DFT PES.
The $\mathcal{AL}$ of the other eight materials, including AgGaS$_{2}$, InNaO$_{2}$, CsBr, CsCl, LiBr, LiCl, LiI, and Na$_{2}$Te, are illustrated in SM~\cite{Supplement} (See Fig.\,S6-S15).
For this purpose, we train an initial MLIP as described in Sec.\,\ref{sec:comp}.
Let us emphasize that no notable artifacts are observed during the training procedure and that further augmenting the training set with data points from the initial aiMD trajectory does not further improve the initial MLIP, see Fig.\,S16 in the SM~\cite{Supplement}.
Subsequent MD runs with this initial MLIP predict the occurrence of strongly anharmonic effects associated with spontaneous defect creation, see Fig.~\ref{fig:AgGaSe2_aiMD}, which displays a MLIP-MD example for AgGaSe$_{2}$ at 300\,K.
Here, the defect creation is observed at 480~ps, as indicated by a jump in the degree of anharmonicity ($A$).

\begin{figure}
\includegraphics[width=0.92\columnwidth]{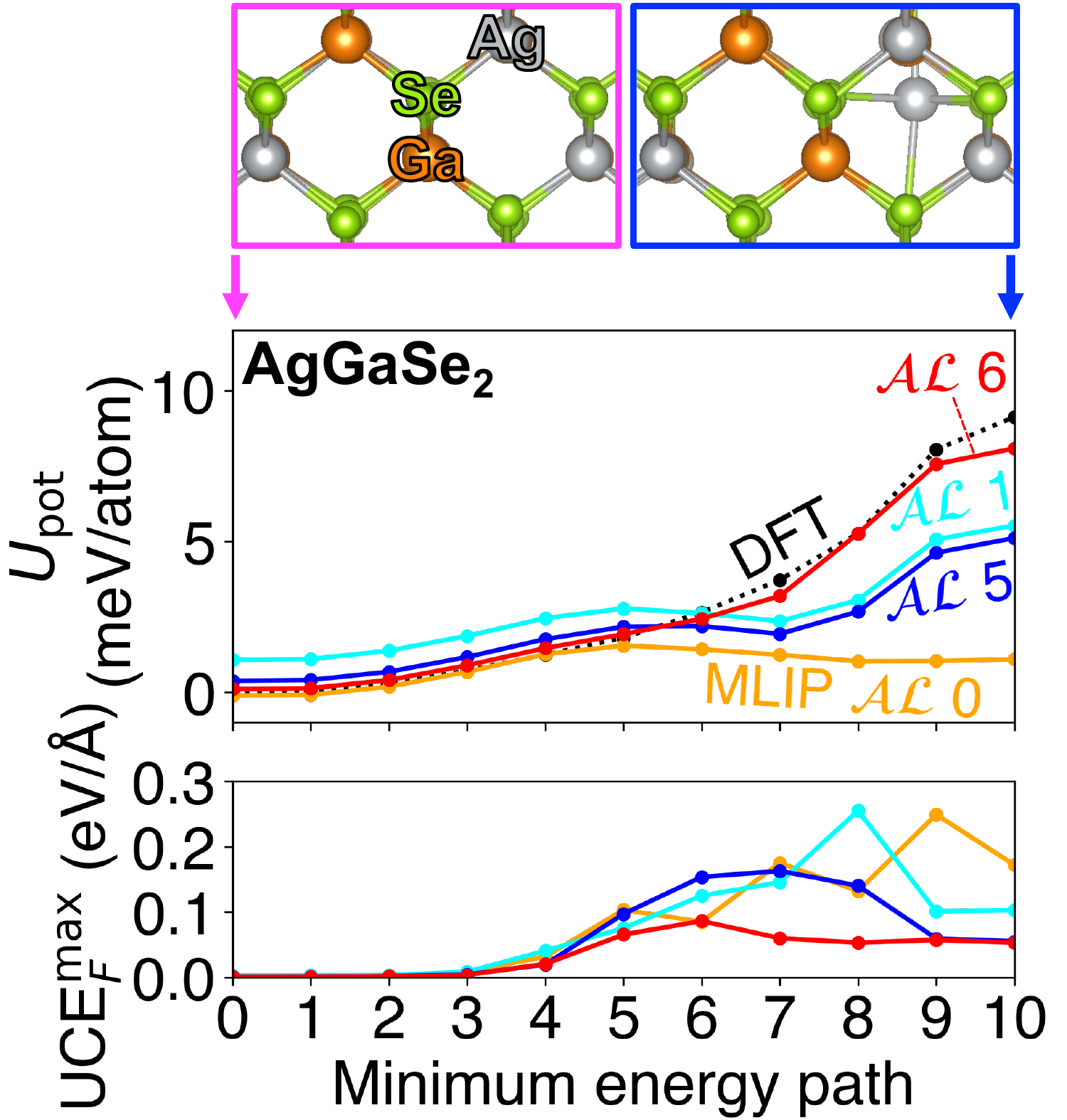}
\caption{\label{fig:AgGaSe2}
The PES and corresponding maximum uncertainty estimates of atomic forces ($\mathrm{UCE}^{\mathrm{max}}_{F}$) predicted by MLIP from $\mathcal{AL}$ using $\mathrm{UCE}^{\mathrm{max}}_{F}$ as a function of the gradual structural evolution of AgGaSe$_{2}$, transitioning from its ground state structure to a defect-bearing configuration via NEB calculations.
}
\end{figure}

To better rationalize the impact of the $\mathcal{AL}$ scheme, we again inspect the minimum-energy path between the pristine
and the defective structures, whereby the latter is obtained from the initial MLIP potential.
As shown in Fig.\,\ref{fig:AgGaSe2} the initial MLIP ($\mathcal{AL}$ 0) predicts a wrong, very likely appeared metastable state, which is not at all present in the ground-truth DFT PES, and there it has a very unfavorable energy.
Again, the devised $\mathcal{AL}$ scheme iteratively improves on the prediction and corrects the erroneous topology of the PES.
These rectifications stem from the sampling of MLIP-MD trajectories traveling beyond the energy barrier during the $\mathcal{AL}$ scheme, featured by the jumps of the degree of anharmonicity in MLIP-MD trajectories illustrated as Fig.~S16.
Whenever unfamiliar events occur during MLIP-MD, our data-sampling method effectively samples these configurational snapshots as sequential training data, resulting in the improvement of MLIP description.
However, since the occurrence of rare events during MLIP-MD is based on chance, corrections of erroneous predictions for such events do not happen in each $\mathcal{AL}$ step, but only in those steps in which such a dynamics is actually observed in Fig.\,\ref{fig:AgGaSe2}.
The uncertainty estimates of maximum atomic force ($\mathrm{UCE}_{F}^{\mathrm{max}}$) become smaller after the sixth step but are still not uniform across the PES, implying that there are still some uncertainty estimates in the high-energy region. 

\begin{figure}
\includegraphics[width=0.875\columnwidth]{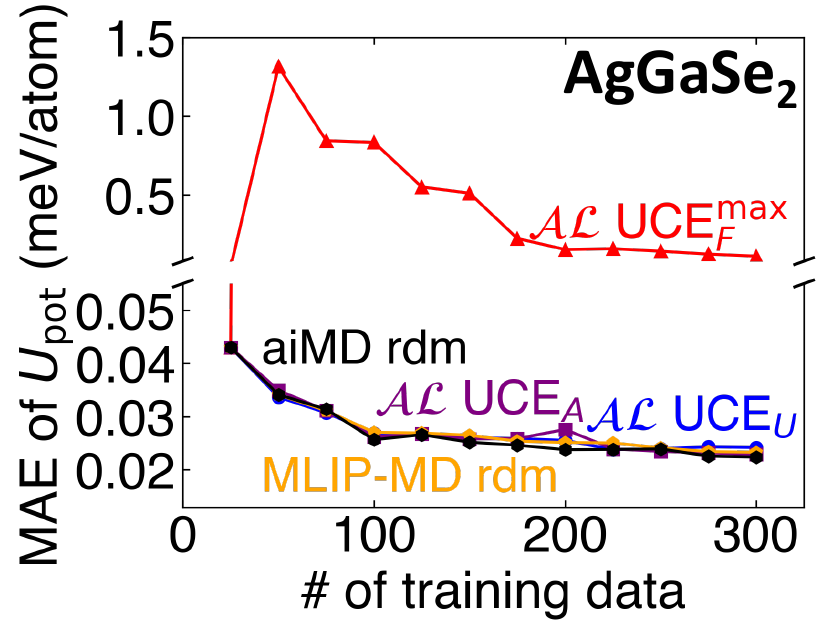}
\caption{\label{fig:AgGaSe2_MAE}
MAE of potential energies in testing data from aiMD trajectory of AgGaSe$_{2}$ as a function of the number of training data.
\texttt{aiMD rdm}~(black hexagon) involves MLIP training with random sampling from aiMD trajectories.
\texttt{MLIP-MD rdm}~(orange pentagon) represents MLIP training with random sampling from MD trajectories using MLIP at each step.
$\mathcal{AL}$~$\mathrm{UCE}_{U}$~(blue circle), $\mathcal{AL}$~$\mathrm{UCE}^{\mathrm{max}}_{F}$~(red triangle), and $\mathcal{AL}$~$\mathrm{UCE}_{A}$~(purple square) exhibit MLIP training with the active learning approach utilizing energy uncertainty estimates, maximum uncertainty estimates of atomic forces, and uncertainty estimates in the degree of anharmonicity, respectively.
}
\end{figure}

Figure \ref{fig:AgGaSe2_MAE} illustrates the MAE  and its converging behavior for potential energies in testing data with different approaches.
\texttt{aiMD rdm} and \texttt{MLIP-MD rdm} exhibit the typical convergence behavior with excellent accuracy for testing data because MLIP predictions for vibrational motions in structure at equilibrium get improved with an increased number of training data.
However, the MLIP models \texttt{aiMD rdm} and \texttt{MLIP-MD rdm}  trained with 300 data still suffer from the occurrence of erroneous, incorrect events in the MLIP-MD simulation (Fig.\,S18~\cite{Supplement} for the \texttt{aiMD rdm} cases).
In addition, the $\mathcal{AL}$ using uncertainties in the energy ($\mathrm{UCE}_{U}$) and the degree of anharmonicity ($\mathrm{UCE}_{A}$) exhibits similar converging behavior compared to \texttt{aiMD rdm}.
This is because $\mathcal{AL}$ $\mathrm{UCE}_{U}$ and $\mathcal{AL}$ $\mathrm{UCE}_{A}$ did not experience any incorrect events during its exploration steps, resulting in no PES corrections in this $\mathcal{AL}$.
This implies that the respective MLIP-MD may likely undergo false events, emphasizing the importance of their appearance during MLIP-MD explorations in $\mathcal{AL}$.
On the other hand, $\mathcal{AL}$ $\mathrm{UCE}_{F}^{\mathrm{max}}$ have a large jump in the MAE of potential energy.
The correction of predictions for the region far from the ground state worsens the prediction for the overall PES.
This deterioration stems from poor regressions due to insufficient training points in the far regions.
Additional sampling is required in these regions, which is not easily conducted due to their high potential energy.
However, this correction is crucial to prevent erroneous events in MLIP-MD, which can seriously affect the dynamic properties of materials, even when happening seldomly.
As long as such incorrect events are prevented, the MAE of below 0.5 meV/atom is still acceptable for applications.

\subsection{\label{sec:physics}Physical Implications of the AL scheme for Practical Simulations}
In the two scenarios above, we have analyzed the effectiveness of the proposed $\mathcal{AL}$ scheme to correct for the erroneously predicted absence and presence of metastable states that induce strongly anharmonic effects.
To judge the overall stability of this approach, we will now consider further long-term MLIP-MD runs.
For the AgGaSe$_2$ case, we ran 30 independent MLIP-MD simulations for up to 1\,ns, as depicted in Fig.\,S19~\cite{Supplement}, and no strongly anharmonic effects could be detected anymore.
Similarly, 30 independent 1\,ns MLIP-MD simulations of CuI from our $\mathcal{AL}$ scheme exhibit defect creations observed in the aiMD trajectory with the correct probabilities and lifetimes.
For the latter, the aiMD trajectory was too short to determine the actual frequency of defect creations.
Instead, we set a pretrained MLIP model as a reference, i.e. the ground truth, and conducted a $\mathcal{AL}$ scheme, yielding new MLIP models.
50 independent 1\,ns MLIP-MD simulations with the new model and the reference model demonstrate that the MLIP models from our $\mathcal{AL}$ scheme can capture the reliable dynamics for these strongly anharmonic events.

\begin{figure}
\includegraphics[width=0.85\columnwidth]{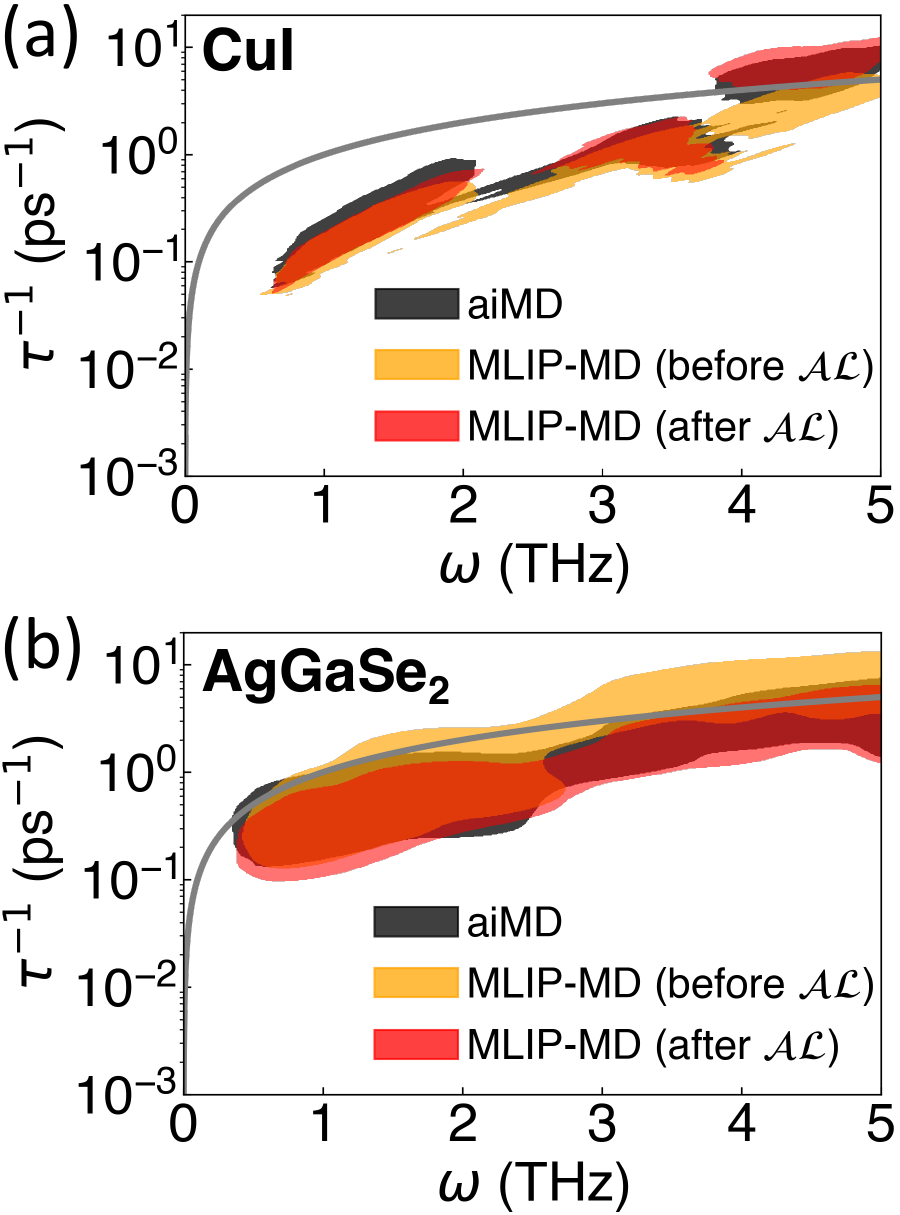}
\caption{\label{fig:md}
The inverse phonon lifetimes population distribution($\tau^{-1}$) as a function of phonon frequency of (a) CuI and (b) AgGaSe$_{2}$, computed by aiMD (Ref.~\cite{Knoop:2023v1}) and MD using MLIPs (MLIP-MD) before/after applying the $\mathcal{AL}$ scheme.
Shaded areas are illustrated for comparison purposes, and their actual, individual point distributions are plotted in Fig \,S20.
These areas are determined by population distribution with Gaussian convolution using a bandwidth of 0.1, and normalized distributions of more than 2\,\% of their maximum populations are shown for each case.
The gray line shows the Ioffe-Regel limit, i.e. a phonon lifetime that corresponds to just one single oscillation.
}
\end{figure}

Further insights can be gained by analyzing the phonon lifetimes obtained from MLIP MD via the fully anharmonic procedure described in Ref.~\cite{Knoop:2023v2}.
As shown in Fig.\,\ref{fig:md} (a), the phonon lifetimes of CuI extracted from the first-principles MD and MLIP MD after $\mathcal{AL}$ are in excellent agreement with each other.
However, the MLIP before $\mathcal{AL}$ significantly overestimates the phonon lifetimes.
 Not surprisingly, AgGaSe$_{2}$ exhibits an inverse effect, in which the lifetimes are underestimated before $\mathcal{AL}$ as illustrated in Fig.\,\ref{fig:md} (b).
Clearly, these trends are related to the fact that the respective MLIPs before $\mathcal{AL}$ either erroneously miss a metastable state or predict one that is actually not there.
In turn, this misses or fictitiously induces strongly anharmonic effects that increase viz. lower the lifetimes, respectively, for CuI and AgGaSe$_{2}$.
Let us emphasize that this change in lifetimes does not depend on whether a spontaneous defect creation is actually observed during the MD trajectory used for the lifetime extraction or not.
The sheer presence of additional minimum results in incessant attempts to overcome the barrier and reach this metastable state.
Even if unsuccessful, these continuous attempts to induce strongly anharmonic effects to massively lower the lifetime.
Notably, this even induces a qualitative change: For CuI, the lifetimes before $\mathcal{AL}$ are low, but still larger than the Ioffe-Regel limit, as plotted in Fig.\,\ref{fig:md} (a).
In DFT and after $\mathcal{AL}$, the strongly anharmonic effects induce a massive reduction of the lifetimes beyond the Ioffe-Regel limit.
Since the phonon picture can break down already when one is close to this Ioffe-Regel limit, it implies that the quasi-particle picture is no longer valid~\cite{Simoncelli:2022}.
As demonstrated in Ref.~\cite{Knoop:2023v1}, this results in a strong reduction of the thermal conductivity that is only accessible with fully anharmonic MD simulations, but not with perturbative phonon-based transport equations.
For AgGaSe$_{2}$, fictitious anharmonic effects are induced due to the erroneous minima in MLIP before $\mathcal{AL}$, inducing the significant underestimation of phonon lifetime as displayed in Fig.\,\ref{fig:md} (b).
This wrong description is corrected in MLIP after $\mathcal{AL}$, closely reproducing the phonon lifetimes from the aiMD trajectory.
In turn, this demonstrates that $\mathcal{AL}$ is absolutely necessary for this system not just for quantitative reasons, but even just to predict the correct qualitative transport regime.

\begin{figure}
\includegraphics[width=0.85\columnwidth]{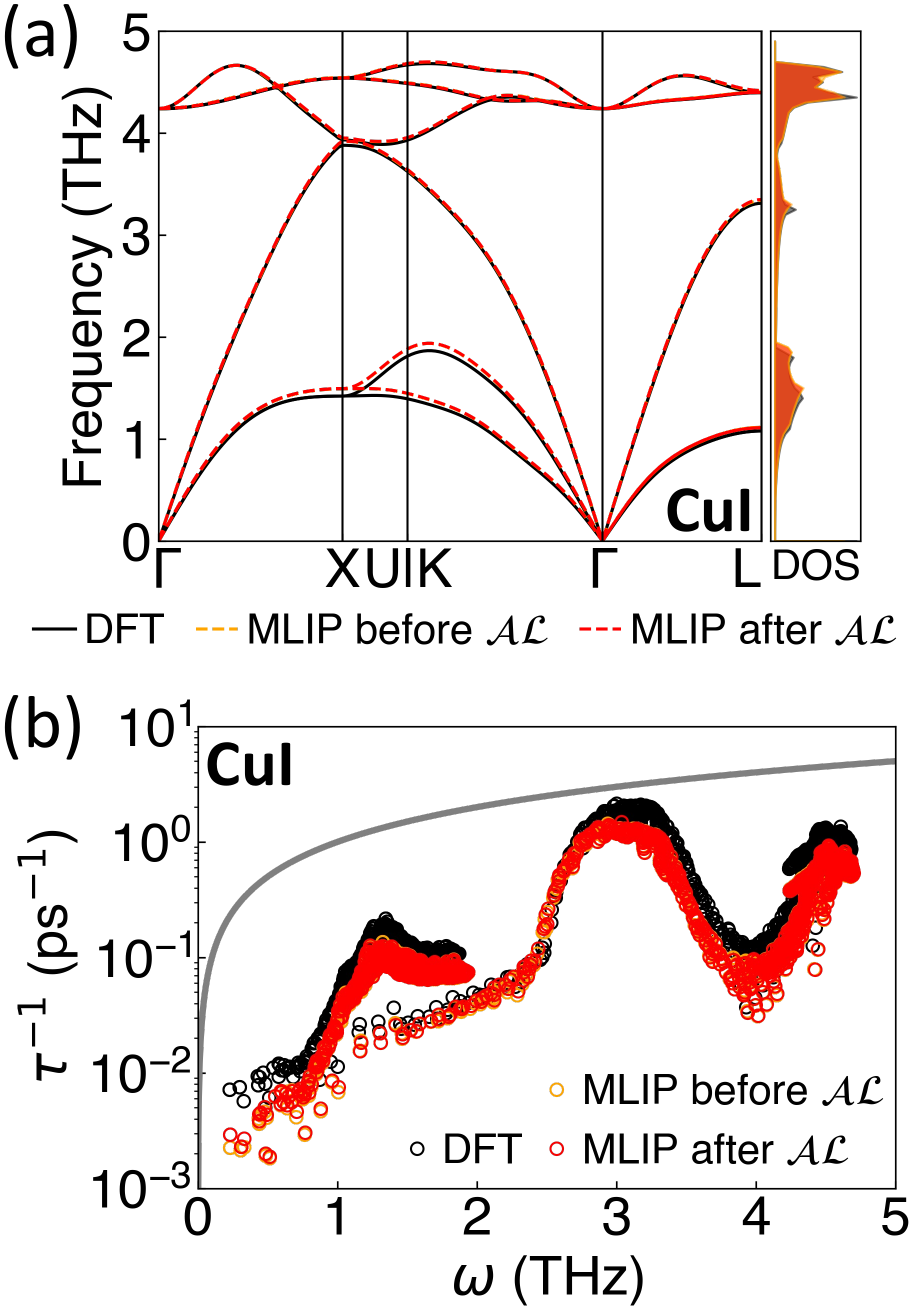}
\caption{\label{fig:phonon}
Perturbative treatment of (a) Phonon dispersion and density of state (DOS) of CuI and (b) inverse phonon lifetimes ($\tau^{-1}$) as a function of phonon frequency of CuI (compare to the non-perturbative results of Fig.\,\ref{fig:md} (a)). The gray solid line in (b) is the Ioffe-Regel limit. The phonon lifetimes of MLIP before and after $\mathcal{AL}$ (Yellow and red markers) in (b) are almost overlapping.
}
\end{figure}

In this context, let us emphasize the importance of using fully anharmonic phonon lifetimes extracted from MD as a metric for judging the reliability of the MLIP.
In the literature, this metric is seldom used because a one-to-one comparison of DFT and MLIP data would require extensive aiMD runs with significant computational costs.
Instead, perturbative techniques for calculating phonon band structures and lifetimes are often used to validate the MLIP against DFT reference data.
However, these techniques only probe the near-equilibrium portion of the PES and can thus lead to erroneously confident conclusions.

For instance, the harmonic phonon band structures obtained properties via perturbation theory are always in excellent agreement between DFT and MLIP, even before $\mathcal{AL}$, as shown in Fig.~\ref{fig:phonon} for CuI and Fig.\,S21~\cite{Supplement} for AgGaSe$_{2}$.
Similarly, phonon lifetimes calculated via perturbation theory from third-order force constant~\cite{Togo:2015, Togo:2023v1} are fairly close between DFT and MLIP, regardless of $\mathcal{AL}$.
This results from the fact that perturbation theory is ``short-sighted'',~i.e.,~it only probes small displacements from equilibrium, but not the full PES accessible in thermodynamic equilibrium.
Accordingly, such techniques are largely insensitive to the presence of additional, metastable minima and to the associated occurrence of strongly anharmonic effects.
This can be further rationalized and substantiated by comparing the perturbative lifetimes in Fig.~\ref{fig:phonon} and Fig.\,S21~\cite{Supplement} to the fully anharmonic ones in Fig.~\ref{fig:md}, which exhibit massively different qualitative and quantitative behavior.
In particular, the perturbative lifetimes of CuI lie mostly well below the Ioffe-Regel limit as shown in Fig.\,\ref{fig:phonon} (b), in sharp contrast to the fully anharmonic ones as illustrated in Fig.\,\ref{fig:md} (b).
With that, perturbative approaches tend to severely underestimate anharmonicity and so to serve as ``self-fulfilling prophecy'', since only the short-range equilibrium dynamics is probed.

\section{\label{sec:conclusions}Conclusions}
In this work, we adapted existing concepts from literature to develop and test an $\mathcal{AL}$ scheme that is suited to train MLIPs that consistently capture strongly anharmonic effects, even if these occur very rarely and are not present in or regularized away from the initial training data. 
Our benchmark on available literature data reveals that, at room temperature, the proposed approach is decisive for 10 out of 112 materials.
In these problematic cases, standard MLIP training procedures either erroneously predict the absence of strongly anharmonic effects or erroneously predict fictitious strongly anharmonic effects.
Using CuI and AgGaSe$_{2}$ as examples, we show that this is related to the presence or absence of meta-stable configurations on the PES that are only seldom explored.
Despite not providing quantitative errors, uncertainty estimates allow to qualitatively detect such problems and can serve as a  warning when the MLIP-MD probes regions uncharted in the training data. 
By exploiting that, the proposed $\mathcal{AL}$ scheme iteratively includes more and more data associated with the phase-space regions featuring problematic predictions, even if those areas are not easily thermodynamically accessible from the start.
With that, the $\mathcal{AL}$ scheme is able to train more precise and accurate MLIPs that correctly account for strongly anharmonic effects with modest computational overhead.
Obviously, the resulting MLIP is not universally valid for the whole structural and thermodynamic phase space. Rather, the proposed $\mathcal{AL}$ scheme can and should be applied whenever different systems and/or thermodynamic conditions are explored,~e.g.,~at new temperatures or pressures, when introducing impurities or defects, under stress or strain, and when introducing new interfaces.

If MLIP models are utilized for systems away from the trained regions, uncertainty estimates must be implemented during MLIP-MD for slightly different systems, e.g., at new temperatures, with impurity atoms, at new interfaces, or under strain.
They serve as the warning alarm to detect when it goes beyond the trained area.

From a physical point of view, our analysis reveals that the proposed $\mathcal{AL}$ procedure is able to produce reliable MLIPs that accurately predict strongly anharmonic effects {\bf without} prior knowledge about the actuating mechanism or about if strongly anharmonic effects are active at all.
In fact, our study also reveals that usual metrics used to monitor the accuracy of MLIPs during training are actually not sensitive to strongly anharmonic effects: Average quantities, like MAE of energies and forces or thermodynamic equilibrium averages, are typically insensitive to such strong anharmonic effects that can be short-lived and occur rarely.
Similarly, standard vibrational properties like phonon frequencies and lifetimes obtained from perturbational ansatzes only probe the near-to-equilibrium region and are, hence, blind to strongly anharmonic effects.
As demonstrated by computing fully anharmonic phonon lifetimes from MD, such approaches are overconfident when it comes to anharmonicity.
This results not just in quantitative errors, but even in the wrong qualitative transport picture.
Conversely, the devised $\mathcal{AL}$ approach reliably accounts for these effects and is able to correctly reproduce all transport regimes.

\begin{acknowledgements}
This project was supported by the ERC Advanced Grant TEC1p (European Research Council, Grant Agreement No. 740233). This research used the resources of the Max Planck Computing and Data Facility.
\end{acknowledgements}

%
%

\end{document}